\documentclass[nofootinbib,twocolumn,aps,prd,amsmath,superscriptaddress]{revtex4-1}
\usepackage{setspace}
\usepackage{color}
\usepackage{fancyhdr}
\usepackage{graphicx}
\usepackage[ansinew]{inputenc}
\usepackage{amssymb}
\usepackage{amsmath}
\usepackage{cancel}
\usepackage{float}
\usepackage{subfigure}
\usepackage{cancel}
\usepackage{tabularx}
\begin{document}
\title{Supermassive Neutron Stars Rule Out  Twin Stars}

\author{Jan-Erik Christian}
\email{christian@astro.uni-frankfurt.de}
\affiliation{Institut f\"ur Theoretische Physik, Goethe Universit\"at Frankfurt, Max von Laue Strasse 1, D-60438 Frankfurt, Germany}
\author{J\"urgen Schaffner-Bielich}
\email{schaffner@astro.uni-frankfurt.de}
\affiliation{Institut f\"ur Theoretische Physik, Goethe Universit\"at Frankfurt, Max von Laue Strasse 1, D-60438 Frankfurt, Germany}
\begin{abstract}
	We investigate the implications of a hypothetical $2.5\mathrm{M_\odot}$ neutron star in regard to the possibility of a first order phase transition to quark matter. We use equations of state (EoS) of varying stiffness provided by a parameterizable relativistic mean field model transitioning to a constant speed of sound EoS. We find a strong connection between the discontinuity in energy density and the maximal neutron star mass. We demonstrate, that high maximal masses cannot be realized for large discontinuities. As a result visible mass gaps, i.e. twin stars, and masses of $M_{max}\gtrsim2.2M_\odot$ are mutually exclusive.
\end{abstract}
\maketitle
\section{Introduction}\label{incanus}
The global properties of neutron stars are uniquely related to the bulk properties of the matter inside, which is given by the equation of state of nuclear matter. The fundamental theory of strong interactions, quantum chromodynamics QCD, can not be solved at present at the high densities encountered inside neutron stars. So one has to resort to observations and detailed modeling  to learn more about the equation of state at high densities.\\
There are currently three main methods of constraining the equation of state of neutron stars. Most commonly used is the mass constraint of about $2\,M_\odot$ \cite{Demorest:2010bx,Antoniadis:2013pzd,Fonseca:2016tux,Cromartie:2019kug,Nieder:2020yqy}, which is the mass of the most massive pulsars measured today. Every viable EoS has to be able to generate a maximal mass higher than this constraint. The radius of a neutron star can be used as a constraint for the EoS as well. NICER \cite{Miller:2019cac,Riley:2019yda,Raaijmakers:2019qny} provides a comparatively precise radius measurement of the millisecond pulsar PSR J0030+0451. With the increased sensitivity of gravitational wave detectors like LIGO/Virgo the tidal deformability of neutron stars can be used as an additional constraint measured from gravitational waves emitted by a binary neutron star inspiral \cite{TheLIGOScientific:2017qsa,Annala:2017llu,Bauswein:2017vtn,Paschalidis:2017qmb,Most:2018hfd,Koppel:2019pys}. The gravitational event GW170817 points to soft EoSs that feature more compact neutron stars \cite{Paschalidis:2017qmb,Alvarez-Castillo:2018pve,Christian:2018jyd,Montana:2018bkb,Sieniawska:2018zzj,Christian:2019qer}. Recently, in the gravitational wave event GW190814 a merger of a black hole with an unknown $2.59^{+0.08}_{-0.09}\,M_\odot$ compact object was observed \cite{Abbott_2020}. This observation has sparked a debate on the maximum mass of neutron stars and the possibility that this unknown object might be a neutron star \cite{Tsokaros:2020hli,Most:2020bba,Godzieba:2020tjn,Fattoyev:2020cws,Dexheimer:2020rlp,Lim:2020zvx,Tews:2020ylw,Tan:2020ics,Zhang:2020zsc,Nunes:2020cuz,Nathanail:2021tay,Blaschke:2020vuy}. In light of this discovery we  investigate the implications of a hypothetical neutron star with such a mass for the possible presence of a phase transition to quark matter.\\
A common type of EoS to describe neutron stars is the relativistic mean field model \cite{PhysRev.98.783,Duerr56,Walecka74,Boguta:1977xi,Serot:1984ey,Mueller:1996pm,Typel:2009sy,Hornick:2018kfi}. Due to the high pressures at the center of a neutron star a quark matter core with a hadronic crust instead of a purely hadronic EoS could be present. This configuration is called a hybrid star \cite{Ivanenko:1965dg,Itoh:1970uw,Alford:2004pf,Coelho:2010fv,Chen:2011my,Masuda:2012kf,Yasutake:2014oxa,Zacchi:2015oma}. The phase transition from hadronic matter to quark matter can lead to a discontinuity in the mass-radius relation. This gives rise to the phenomenon known as twin stars, where two neutron stars have the same mass, but different radii \cite{Kampfer:1981yr,Glendenning:1998ag,Schertler:2000xq,SchaffnerBielich:2002ki,Zdunik:2012dj,Alford:2015dpa,Blaschke:2015uva,Zacchi:2016tjw,Alford:2017qgh,Christian:2017jni,Blaschke:2019tbh,Jakobus:2020nxw}. Hybrid and twin stars tend to be rather compact, which is in good agreement with GW170817 \cite{Paschalidis:2017qmb,Alvarez-Castillo:2018pve,Christian:2018jyd,Montana:2018bkb,Sieniawska:2018zzj,Christian:2019qer}. However, it implies a soft hadronic EoS, which can come into conflict with the mass constraint. In this work we will use the parameterizable relativistic mean field equation of state in the form presented by Hornick et al. \cite{Hornick:2018kfi} and combine it with a constant speed of sound (CSS) approach for quark matter \cite{Zdunik:2005kh,Alford:2014dva,Alford:2015gna} via a Maxwell-construction. This ansatz gives us the opportunity to vary the parameters and apply the hypothetical GW190814 constraint in a more generalized way. We also discuss the $2.14^{+0.10}_{-0.09}\,M_\odot$ pulsar mass constraint from Cromartie et al. \cite{Cromartie:2019kug}. 
We find, that EoSs featuring twin stars and masses of $M_{max}\gtrsim2.2\,M_\odot$ are mutually exclusive. However we note that hybrid stars without a discontinuity in the mass radius relation are still viable and can even reach maximal masses of  $M_{max} > 2.5\,M_\odot$, as discussed in \cite{Tan:2020ics}.
\section{Theoretical Framework}
\subsection{Equation of State} 
\subsubsection{Hadronic Equation of State} 
The EoS we use here was  introduced by Todd-Rudel et al. \cite{ToddRutel:2005fa} (see also: Chen et al. \cite{Chen:2014sca}) and is a generalized relativistic mean field approach with the main advantage, that the slope parameter $L$, the symmetry energy $J$ and the effective nucleon mass $m^*/m$ can be easily adjusted.  Hornick et al. \cite{Hornick:2018kfi} additionally constrain the parameters using the constraints from  $\chi\mathrm{EFT}$ for densities up to $1.3\,n_0$ \cite{Drischler:2016djf}.\\
The choices of $L$ and $J$ have no significant impact on the mass radius relation as shown in ref. \cite{Hornick:2018kfi}.This allows us to fix the values $L=60\,\mathrm{MeV}$ and $J=32\,\mathrm{MeV}$ in a way, that provides the largest range in $m^*/m$ from  $m^{*}/m=0.55$ to $m^{*}/m=0.75$. Note that the stiffness of an EoS relates to the value of $m^{*}/m$  \cite{1983PhLB..120..289B}. The lower the effective mass parameter, the stiffer is the EoS.
\subsubsection{Phase Transition}
We consider a first order phase transition at high baryonic densities from hadronic to quark matter EoS. 
 For the hadronic matter we use the parameterized EoS (see previous section), while the constant speed of sound approach \cite{Zdunik:2012dj,Alford:2014dva,Alford:2015gna} is employed for the quark matter. This means, the entire EoS is given as:
\begin{equation}
\epsilon(p) =
\begin{cases} 
\epsilon_{HM}(p)	&  p < p_{trans}\\
\epsilon_{HM}(p_{trans})+\Delta\epsilon + c_{QM}^{-2}(p-p_{trans})	& p > p_{trans}\\
\end{cases}
\end{equation} 
where $p_{trans}$ is the transitional pressure and $\epsilon_{HM}(p_{trans})$ the energy density at the point of transition.
The discontinuity in energy density at the transition is $\Delta\epsilon$. 
In order to achieve the stiffest possible EoSs and thus the greatest possible range of mass-radius relations we set $c_{QM}=1$ using natural units. Lower values of $c_{QM} $ can not generate higher maximal masses, if the other parameters are identical and $c_{QM}=1$ allows for the greatest range of parameters.
\subsection{Classification of Twin Stars}\label{Tevildo}
\begin{figure}
	\centering				
\includegraphics[width=8.6cm]{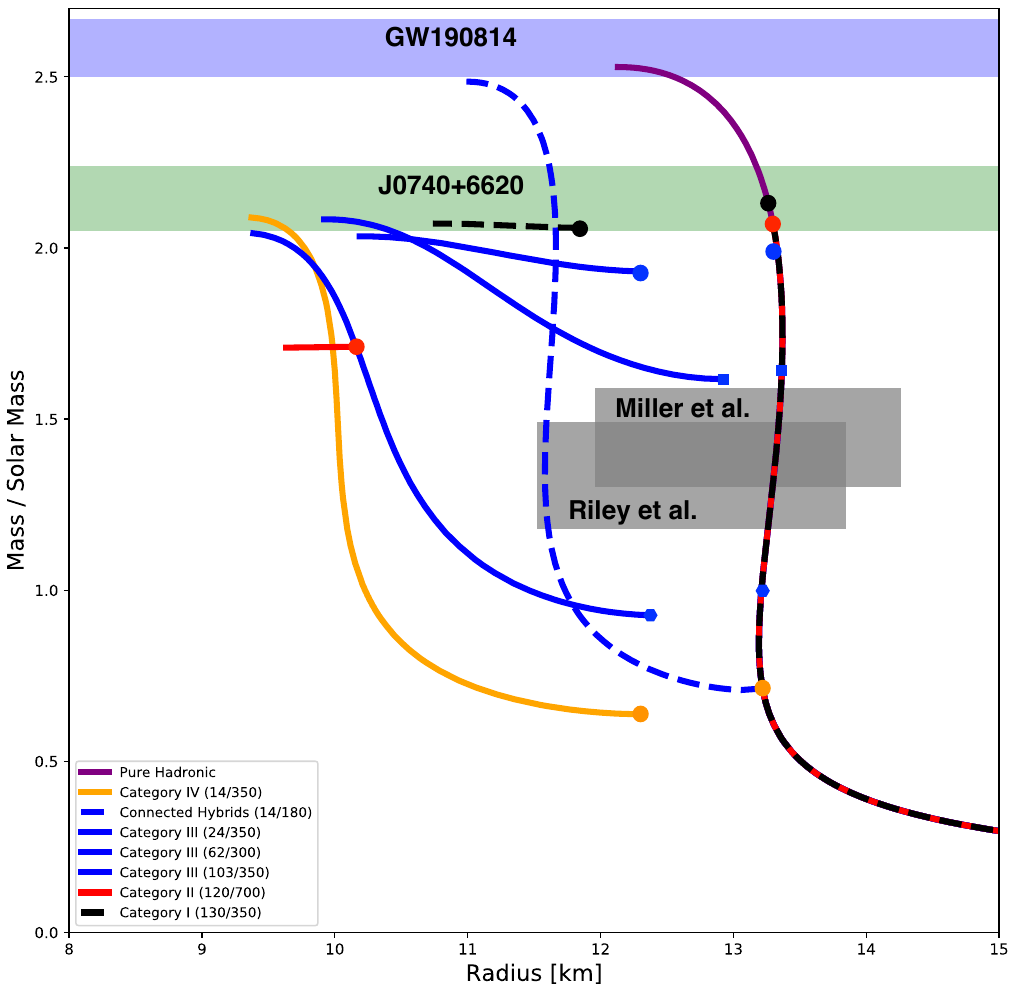}
\caption{\footnotesize Examples for category I (black dashed line), II (red dashed line), III (blue continuous lines) and IV (orange line) phase transitions for a hadronic base EoS with $m^*/m = 0.65$. As well as the pure hadronic case (purple line) and an example of a connected second branch (dashed blue line). The values in parenthesis indicate the transitional pressure and the jump in energy density in units of $\mathrm{MeV/fm^3}$ respectively. The point, where the first branch ends is 
indicated by a dot in the color corresponding to the category. Additionally the NICER constraints \cite{Miller:2019cac,Riley:2019yda} (black shaded area), pulsar mass limit \cite{Cromartie:2019kug} (green shaded area) and the mass of the unknown compact object of GW190814 \cite{Abbott_2020} (blue shaded area) were added.}
\label{Cats}
\end{figure}
A first order phase transition can give rise to the phenomenon of "twin stars", which are neutron stars with identical mass, but different radii  \cite{Kampfer:1981yr,Glendenning:1998ag,Schertler:2000xq,SchaffnerBielich:2002ki,Zdunik:2012dj,Alford:2015dpa,Blaschke:2015uva,Zacchi:2016tjw,Christian:2017jni}. When analyzing such EoSs it can be useful to classify the twin star solutions into four distinct categories, as described in \cite{Christian:2017jni}. We refer to the maximum of the hadronic branch as the first maximum $M_1$ and the maximum of the hybrid branch as the second maximum $M_2$ in a twin star mass-radius relation. The mass value of the first and second maximum can be related to values of $p_{trans}$ and $\Delta\epsilon$ respectively \cite{Christian:2017jni}. The shape of the second branch is correlated with the value of $p_{trans}$, while its position is strongly influenced by the value of $\Delta\epsilon$. High values of $p_{trans}$ lead to high masses in the first maximum and flat second branches. Low values of $\Delta\epsilon$ lead to a second branch near the discontinuity (i.e. a high mass at the second maximum).
Based on this observation the categories can be defined as:
\begin{itemize}
	\item [\textbf{I:}] Both maxima exceed $2\,M_\odot$.
	\item [\textbf{II:}] Only the first maximum reaches $2\,M_\odot$.
	\item [\textbf{III:}]The first maximum is in the range of $2\,M_\odot \geq M_{max_1} \geq 1\,M_\odot$, while the second maximum exceeds $2\,M_\odot$. 
	\item [\textbf{IV:}] Like category III  the second maximum exceeds $2\,M_\odot$, however the first maximum is below even $1\,M_\odot$.
\end{itemize}
Due to the quark matter dominated EoSs in category IV the hadronic part can be nearly arbitrarily soft and the combination can still reach $2\,M_\odot$. Category I-III do not allow for extremely soft nuclear EOSs with effective masses of $m^{*}/m\ge0.75$.\\
Examples for all four categories are provided in Fig. \ref{Cats}. The base hadronic EoS in this example is the parameterized relativistic mean field model with an effective mass of $m^{*}/m=0.65$. Category I (dashed black line) and category II (red line) have their hadronic maxima close to each other. In general all values of $p_{trans}$ in category II are also included in category I. The distinguishing factor is the higher jump in energy density for the category II cases, which moves the flat second branch further down in the mass-radius diagram.\\
Since category III contains the greatest variety of  shapes we include three examples for this category (continuous blue lines): One at the lower limit of the transitional pressure, i.e. $M_1 \simeq 1\,M_\odot$ ($p_{trans} = 24\,\mathrm{MeV/fm^3}$), one at the upper limit, i.e. $M_1 \simeq 2\,M_\odot$ ($p_{trans} = 103\,\mathrm{MeV/fm^3}$) and one example from the mid-range ($p_{trans} = 62\,\mathrm{MeV/fm^3}$). With decreasing transitional pressure the second branch becomes much steeper when compared to previous categories.\\
Fig. \ref{Cats} also includes a connected hybrid star configuration with an early phase transition (dashed blue line). A connected branch has a kink at the point of transition and is realized for small values of $\Delta\epsilon$. Hybrid stars with a connected branch can reach higher masses than true twin star configurations, however the presence of a phase transition can be hard to determine from mass-radius measurements. 
It is possible for such a connected hybrid star branch to feature stars of similar masses and significantly different radii, which could also be interpreted as twin stars, as, for example, Blaschke and Cierniak do \cite{Blaschke:2020vuy}. However in the following we will only consider rising twin star cases, i.e. configurations where the more massive twin has a larger radius \cite{Schertler:2000xq}, as these configurations are accompanied by gaps in the mass radius relation, which will ensure a measurable difference in radius for the considered twin star cases.
\section{Trying to reconcile strong phase transitions  with a hypothetical 2.5$M_\odot$ neutron star} 
\subsection{The stiffest case}
\begin{figure}
	\centering				
	\includegraphics[width=8.6cm]{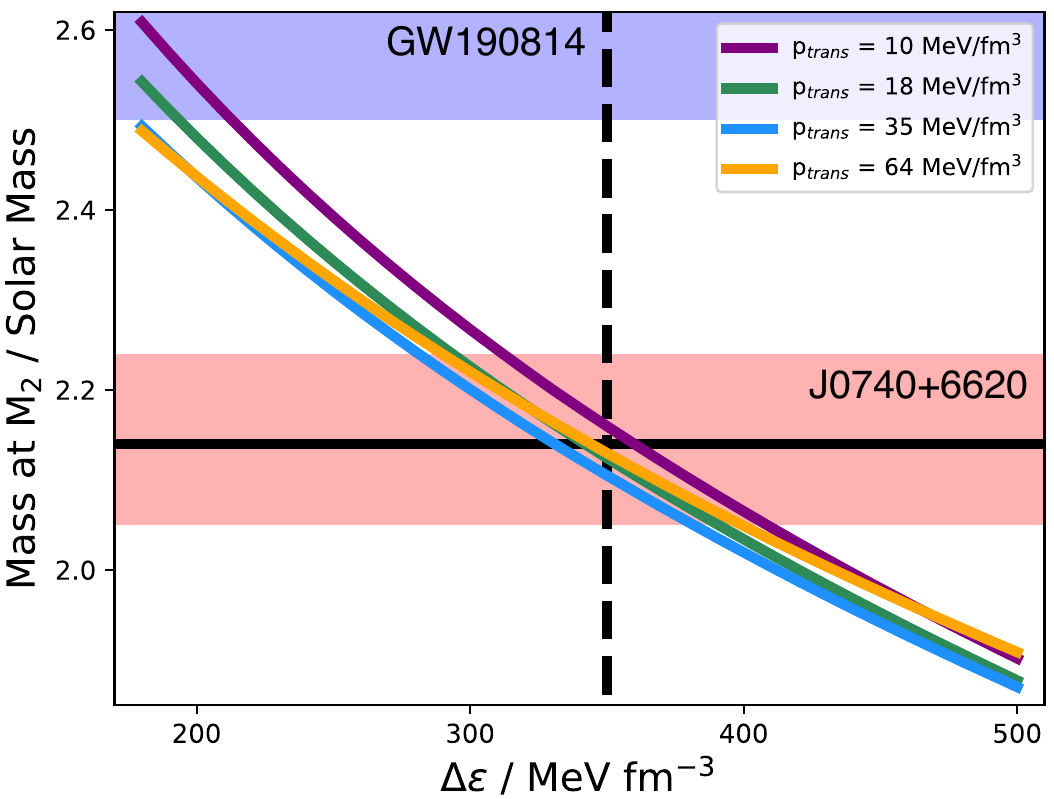}
	\caption{\footnotesize The $m^{*}/m=0.55$ case of the maximal hybrid star mass as a function of the jump in energy density for different transitional pressures. The dashed black line indicates $\Delta\epsilon = 350\,\mathrm{MeV/fm^3}$, which is the lowest value of $\Delta\epsilon$ that generates a mass gap of $\Delta M = \pm0.1\,M_\odot$ for all transitional pressures. The hypothetical neutron star participant of GW190814 (shaded blue) is only realizable for smaller values of $\Delta\epsilon$, implying no viable mass gap. The solid horizontal black line represents the $2.14\,M_\odot$ of J0740+6620 \cite{Cromartie:2019kug}. Only the $p_{trans} = 10\,\mathrm{MeV/fm^3}$ case reaches this value at $\Delta\epsilon\le 350\,\mathrm{MeV/fm^3}$, suggesting, that only this case can support twin stars and a maximal hybrid star mass of $M_{2}\ge2.14\,M_\odot$. However, when considering the error bar of the stated mass (red shaded area) all cases reach the limit.}
	\label{MDE}
\end{figure}
Since the case $m^{*}/m=0.55$ is the stiffest possible hadronic equation in our model it will feature the most massive neutron stars. Therefore any cases excluded by this EoS due to an insufficient maximum mass are excluded for all cases $m^{*}/m>0.55$ as well.\\ 
Fig. \ref{MDE} shows the masses at the hybrid star maximum $M_2$, as they depend on the jump in energy density $\Delta\epsilon$. 
 In our previous publication \cite{Christian:2019qer} we used the NICER measurement \cite{Raaijmakers:2019qny,Miller:2019cac,Riley:2019yda} at the $2\sigma$ level to constrain the minimal density at the core  for strong phase transitions to $n \approx 1.4\,n_0$. This corresponds to about $p_{trans} \ge 10\mathrm{MeV/fm^3}$ for the $m^{*}/m=0.55$ case used here (purple line in Fig. \ref{MDE}). The $1\sigma$ accuracy of the NICER radius constraint \cite{Raaijmakers:2019qny,Miller:2019cac,Riley:2019yda} corresponds to $n\approx1.7\,n_0$ \cite{Christian:2019qer} which is about $p_{trans} \ge 18\mathrm{MeV/fm^3}$ for this EoS (green line in Fig. \ref{MDE}). The cases $p_{trans} = 35\mathrm{MeV/fm^3}$ (blue line) and $p_{trans} = 64\mathrm{MeV/fm^3}$ (orange line) are the middle and upper limit of category III phase transitions respectively. This means that those two cases correspond to maximal masses of  about $2\,M_\odot$ and about $1.5\,M_\odot$ in the first branch.
The $p_{trans} = 64\mathrm{MeV/fm^3}$ is not compact enough to fit with the GW170817 measurement \cite{Abbott:2018wiz}, but was included for completeness.\\
We considered a phase transition to be strong if the jump in mass between the hadronic maximum and the hybrid star minimum is larger than $0.1\,\mathrm{M_\odot}$. This is possible for every EoS in our model, if the discontinuity in energy density is $\Delta\epsilon \ge 350\,\mathrm{MeV/fm^3}$ and a twin star branch is generated at all. The condition $\Delta\epsilon \ge 350\,\mathrm{MeV/fm^3}$ is not a rigorous constraint, but a good approximation, as all cases $\Delta\epsilon = 350\,\mathrm{MeV/fm^3}$ lead to $\Delta M \ge0.1\,\mathrm{M_\odot}$, even if there are cases where a lower jump in energy density would be sufficient. 
The case $p_{trans}=10\mathrm{MeV/fm^3}$ has a maximal mass of about $2.2\,M_\odot$ at a discontinuity of $\Delta\epsilon=350\mathrm{MeV/fm^3}$. 
The most massive neutron stars known today \cite{Cromartie:2019kug} is indicated in Fig. \ref{MDE} as a black vertical line at $2.14\,M_\odot$ with a red shaded error bar of $^{+0.10}_{-0.09}$. All cases shown reach at least the $1\sigma$ error-bar for $\Delta\epsilon \ge 350\,\mathrm{MeV/fm^3}$.\\
 However, masses as large as the hypothetical $2.59^{+0.08}_{-0.09}\,M_\odot$ posed by  GW190814 \cite{Abbott_2020} would imply, that a strong phase transition that produces visible twin stars can not be realized in nature. Hybrid stars with a connected branch (i.e. cases with $\Delta\epsilon \le 350\,\mathrm{MeV/fm^3}$ ) could still fit the data.
Category I and II of the $m^{*}/m=0.55$ case contain maximal masses of, at most, $M_{max}\simeq 2.35\,M_\odot$ with the maximal hybrid star mass being similar or lighter than the purely hadronic maximum. Since category I contains the highest possible maximal mass of all four twin star categories and the $m^{*}/m=0.55$ case contains the most massive stars it follows, that a neutron star measurement of  $M> 2.35\,M_\odot$ would rule out twin stars. However, since the $m^{*}/m=0.55$ case is too stiff to produce stars compact enough to match the GW170817 measurement without an early phase transition \cite{Christian:2019qer}, category I and II configurations of this case are not compatible with this event. This means the category IV case of about $M_{max}\simeq2.2\,M_\odot$ is the highest possible mass a twin star configuration can reach.\\

\subsection{Softer Cases}
\begin{figure}
	\centering				
	\includegraphics[width=8.6cm]{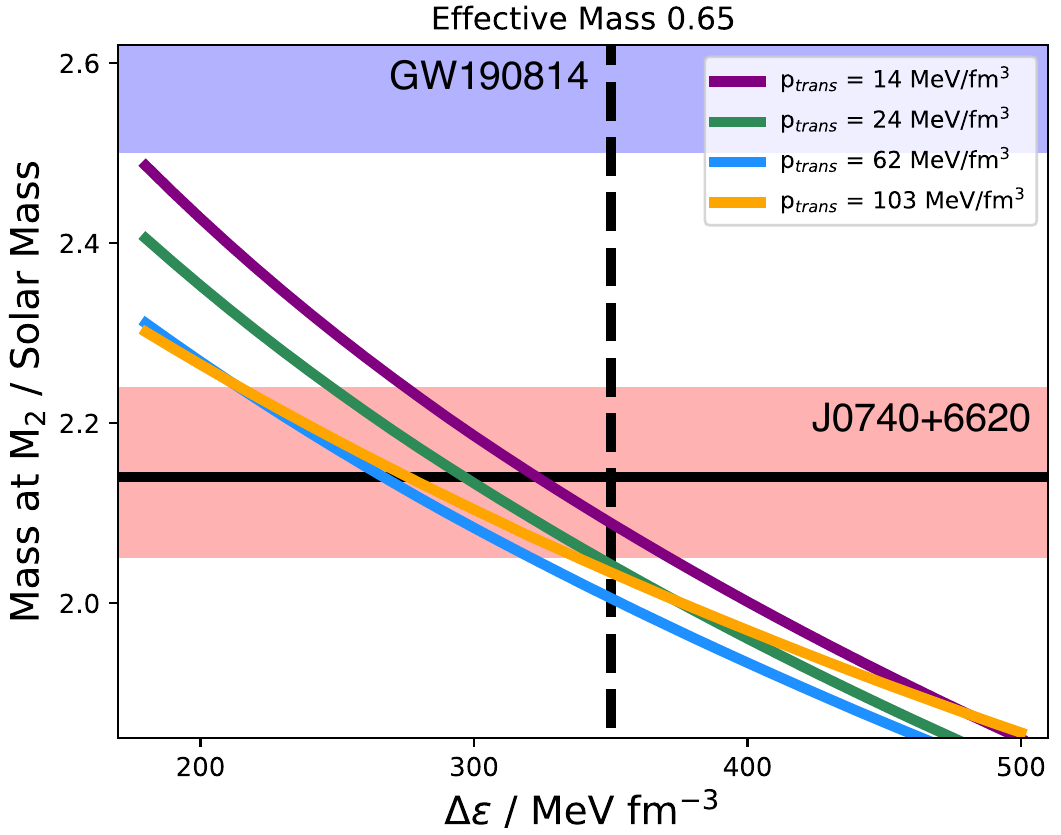}
	\caption{\footnotesize Most transitional pressures of category III in the $m^{*}/m=0.65$ case do not reach the pulsar mass limit of J0740+6620 at $\Delta\epsilon \ge 350\,\mathrm{MeV/fm^3}$. Therefore only the cases where $p_{trans}\le 24\,\mathrm{MeV/fm^3}$ have a mass gap of $0.1\,M_\odot$ and generate a mass value compatible with Cromartie et al. \cite{Cromartie:2019kug} at the $1\sigma$ interval.}
	\label{MDE65}
\end{figure}
In this section we study the parameterizations with higher effective masses, i.e. softer nuclear EoSs.  
We find, that only $m^{*}/m\le0.60$ cases support category III phase transitions that generates (connected) hybrid stars with masses larger than $2.5\,M_\odot$.
The EoS with the highest effective mass still generating visible twin stars  and meeting the $2.14^{+0.10}_{-0.09}\,M_\odot$  mass constraint outside of category IV is the $m^{*}/m=0.65$  case.
In Fig. \ref{MDE65} the purple line represents the transitional pressure corresponding to $1.7n_0$.
The low value of this transitional pressure places the mass-radius relations in category IV. The remaining three cases shown  represent the lower limit (green line), a for the category representative example (blue line) and the upper limit (orange line) of category III respectively.\\
 In comparison with Fig. \ref{MDE} the impact of the transitional pressure on the position of the $M_2$ curve becomes more apparent. Since higher values of the effective mass lead to softer EoSs the transitional pressures corresponding to 1, 1.5 and $2\,M_\odot$ rise, leading to lower $M_2$ curves, that are further apart, than at lower effective masses. The $m^{*}/m=0.65$  EoSs is the stiffest case investigated here, that produces neutron stars compact enough to meet the GW170817 constraint without a phase transition. This means, that the highest possible category I configuration compatible with GW170817 is the $M_{max}=2.15\,M_\odot$ case of the $m^{*}/m=0.65$ EoS.\\\\
We summarize the compatibility of the four categories and the five investigated effective masses in table \ref{Tablenew}. Here the letters a, b, c, d and e stand for the effective masses 0.55, 0.60, 0.65, 0.70 and 0.75 respectively.  The appearance of  a letter in a cell indicates, that the constraint (row) is fulfilled by the related category (column) for the corresponding effective mass. The brackets indicate, that an effective mass only partially or narrowly fulfills the constraint. In the case of the NICER constraint this applies to all category IV phase transitions, since the majority of them would be excluded by the $1\sigma$ constraint of the radius constraint, but not by the $2\sigma$ interval. We note, that all calculations are for nonrotating stars. The maximal mass increase due to rotation is 20\% at the Keplarian limit \cite{Breu:2016ufb}. However, most pulsar do not rotate that fast and the mass correction at lower frequencies is within a few percent.
\begin{table}
	\begin{center}
		\begin{tabular}{|c|c|c|c|c|c|}
			\hline
			\hspace{.2cm} \hspace{.2cm} & \hspace{.01cm} CI
			\hspace{.1cm} & \hspace{.2cm} CII \hspace{.2cm} & \hspace{.1cm}
			CIII \hspace{.1cm} & \hspace{.1cm}
			CIV \hspace{.1cm} & \hspace{.1cm}
			Connected \hspace{.1cm}\\
			\cline{1-6}
			NICER & abcd & abcd & abcd & (abcde) & abcde\\ 
			GW170817 & cd & cd & (a)bcd & abcde  & abcde\\ 
			 $2.05\,M_\odot$ & abcd & abc & abc & abcde & abcde\\ 
			$2.50\,M_\odot$ & / & / & / & / & ab\\
			\hline
		\end{tabular}
		\caption{Shown are effective masses that fulfill the constrains for categories of twin stars. The letters a, b, c, d and e correspond to the effective masses 0.55, 0.60, 0.65, 0.70 and 0.75 respectively. We only show the lower limit of the mass constraints from Cromartie et al. \cite{Cromartie:2019kug} and the unknown compact object of GW190814 \cite{Abbott_2020}.The tidal deformability constraint from GW170817 favors soft EoSs, while the mass constraints favor stiff EoSs. Only the case $m^*/m = 0.65$ fulfills all constraints for all categories, except for GW190814, which can not be fulfilled by any category.}
		\label{Tablenew}
	\end{center}
\end{table}
\section{Conclusion}\label{conclusion}
We combined the parameterizable relativistic mean field equation of state with a constant speed of sound approach for quark matter to make a general parameterized study about the possible existence of  hybrid stars and twin stars. For this investigation we compare the mass-radius relations of EoSs with varying effective nucleon masses $m^{*}/m$ (and thus varying stiffness) with mass constraints from a hypothetical $2.5\,M_\odot$ neutron star and the $2.14^{+0.10}_{-0.09}\,M_\odot$ of the pulsar J0470+6620 \cite{Cromartie:2019kug}. We find that only the cases $m^{*}/m=0.55$ and $m^{*}/m=0.60$ can generate hybrid stars massive enough to reach $2.5\,M_\odot$. However, these configurations would consist of a single mass-radius line with a kink, instead of two seperate branches featuring  twin stars. Effective masses of $m^{*}/m\le0.65$ can generate visible twin stars with a $0.1\,M_\odot$ mass gap that reach the limit of $2.14^{+0.10}_{-0.09}\,M_\odot$ \cite{Cromartie:2019kug}.\\
Furthermore we considered the tidal deformability constraint from the LIGO/Virgo measurement of GW170817 \cite{Abbott:2018wiz}, which favors soft EoSs that generate compact stars \cite{Paschalidis:2017qmb,Alvarez-Castillo:2018pve,Christian:2018jyd,Montana:2018bkb,Sieniawska:2018zzj}. This results in the effective masses $m^{*}/m=0.55$ and $m^{*}/m=0.60$ being disfavored, unless an early phase transition is present (see \cite{Christian:2019qer}). NICERs measurement of J0030+0451 \cite{Raaijmakers:2019qny,Riley:2019yda,Miller:2019cac} suggests a phase transition at masses of  at least $1\,M_\odot$, as most earlier phase transitions usually generate too small radii \cite{Christian:2019qer}.\\
We find, that a mass constraint of  $2.5\,M_\odot$ would rule out all twin star solutions, as well as most hybrid star solutions. The maximal mass that allows for twin star solutions is about $2.2\,M_\odot$, which can be realized for the stiffest EoS in our model at the lowest possible transitional pressure. We also find that for a maximal pulsar mass of $2.15\,M_\odot$ and larger a possible phase transition has to be present below neutron star mass configurations of $2\,M_\odot$. 
We investigated the robustness of our results by decreasing the speed of sound. We find that a speed of sound of at least $c_{QM}=0.8$ is necessary to generate a twin star configuration with a $0.1\,M_\odot$ mass gap and a maximal mass above $2M_\odot$ that is compatible with the constraints from GW170817. Even for those cases only category II configurations are possible, as the quark matter EoS is not stiff enough to reach high masses. The effective mass case of $m^*/m=0.65$ narrowly reaches a maximal masses above $2M_\odot$ and is therefore the only case compatible with GW170817, when a speed of sound $c_{QM}=0.8$ is used.\\
It is likely, that future third generation wave detectors such as the Einstein telescope \cite{Maggiore:2019uih} will be able to detect the tidal deformability of objects like the hypothetical $2.5M_\odot$ hybrid star, which will constrain the EoS significantly more.
\begin{acknowledgments}
 J.E.C. thanks the Giersch foundation for their support with a Carlo and Karin Giersch Scholarship. 
\end{acknowledgments}
\bibliographystyle{apsrev4-1}
\bibliography{neue_bib_JSB}
\end{document}